# On the Complexity of the Cycles-based Synthesis of Ternary Reversible Circuits

Caroline Barbieri [1][0000-0002-9348-7406], and Claudio Moraga [2][0000-0002-4423-8112]

[1] Instituto Federal de Educação, Ciência e Tecnologia de São Paulo Campus Capivari, São Paulo, Brazil
[2] Chair Informatics 1, TU Dortmund University, 44227 Dortmund, Germany
carolbarbieri1@gmail.com, claudio.moraga@udo.edu

**Abstract.** The paper studies the main aspects of the realization of 2×2 ternary reversible circuits based on cycles, considering the results of the realization of all 362,880 2×2 ternary reversible functions. It has been shown that in most cases, realizations obtained with the MMD+ algorithm have a lower complexity (in terms of cost) than realizations based on cycles. In the paper it is shown under which conditions realizations based on transpositions may have a higher or a lower cost than realizations using larger cycles. Finally it is shown that there are a few special cases where realizations based on transpositions have the same cost or possibly lower cost than the MMD+ based realizations. Aspects of scalability are considered in terms of 2×2-based $n \times n$ reversible circuits.

## 1 Introduction

Research devoted to Reversible Computing circuits has possibly been motivated by its capability of reducing power dissipation as heat and because of its closeness to quantum computing [13]. Two happenings of the past decade underline its importance: First, a series of annual meetings on Reversible Computing, starting 2010, as a Workshop and soon after, becoming an annual Conference (see e.g. [5]). Second, a COST-Action IC-1405: *Reversible Computation – Extending Horizons of Computing* [1], supporting a network of researchers mainly from countries of the European Community.

At the hardware level, most works are dedicated to binary reversible circuits, possibly for the compatibility with the prevailing digital world. The fact that this is (mainly) binary is due to the fact that along the years the available basic components, (relays, tubes, transistors), have two reliable, controllable, stable states. In the

The research work of C.B. leading to this paper was supported by a doctoral scholarship - 88881.134365/2016-01 of the Capes Foundation, Ministry of Education, Government of Brazil. The research work of C.M. was partially supported by the EU COST-Action IC1405 on Reversible Computation – Extending Horizons of Computing.

context of quantum computing the above constraint does not exist. Physicists have shown that gates working on more than two values are feasible. Possibly the most relevant work related to this aspect is that of A. Muthukrishnan and C.R. Stroud Jr. [12] who succeeded in building *p*-valued gates using a linear ion trap approach. This opened the door to *p*-valued reversible/quantum circuit developments. The simple case of choosing $p = 3$ shows rather unexpected possibilities. In the binary case, there are $2^3! = 40,320$ reversible functions of *three* variables, whereas in the ternary case there are $3^2! = 362,880$ functions of *two* variables. This is the simplest possible example of the potential for computational power that could be expected in the further development of *p*-valued reversible/quantum circuits. The present paper intends to be a contribution in that direction.

The synthesis of minimal reversible binary circuits is in NP. The main different approaches to this problem use databases of partial solutions [6], or look for a symbolic representation of the problem and apply e.g. SAT solvers [7]. Among approaches that proceed without additional support, the most well known use the transformation based MMD algorithm [10], [17], genetic programming to synthesize reversible circuits [8], or decompose the permutation that specifies a reversible function into cycles. V.V. Shende *et al.* introduced the cycles method for binary reversible circuits in [16], not only decomposing permutations into their natural cycles, but further decomposing cycles –(which are permutations)– into transpositions, which are cycles of length 2. This method was reviewed by Z. Sasanian *et al.* [15], who proved that decomposing the natural cycles into cycles of length 3 in many cases would lead to circuits with a lower number of gates than by decomposing into transpositions. A similar result holds for 2×2 ternary reversible circuits.

In [11] it was shown that the MMD algorithm could be successfully extended to the ternary case. The present study extends the results of [11] by considering realizations based on cycle decompositions. Cycle-based realization of ternary reversible circuits have been reported (see e.g. [18]), but without a cost-comparison with MMD realizations. An evaluation of realizations based on decomposition of permutations into cycles of different lengths using as a benchmark the whole set of $(3^2)! = 362,880$ ternary reversible functions of 2 variables (as in [11]) was first reported in [2] and showed that realizations done with the MMD+ algorithm [3] in 96.7% of the benchmark had cost no higher – and in 54.4% a lower cost – than realizations based on cycles of natural length. (In the MMD+ algorithm the choice of each elementary permutation must satisfy two optimization criteria: an elementary permutation is chosen if it fulfills the intended local transformation and minimizes the distance to the specified global permutation, and if there are several elementary permutations that satisfy this requirement, the one with the lowest realization cost will be chosen. If again several of the appropriate elementary permutations have the same low cost, then a random choice will be made.) Furthermore, in [2] it was also shown that in 69.9% of the benchmark functions, not decomposed cycles have lower realization cost than cycles decomposed in 3-cycles. In most cases realizations based on 3-cycles have lower cost than realizations based on transpositions. In what follows an example will be shown, selected at random from the large set of functions (decomposable into three cycles) to illustrate that the design based on cycles has a higher cost than direct realization of the original permutation, and that decomposition into transitions has an even higher cost. Reasons for the results will be discussed.

## 2  Case analysis

Let F = (4, 3, 7, 5, 8, 1, 2, 6, 0). It is simple to show that F may be expressed as (0, 4, 8)(1, 3, 5)(2, 7, 6). Since the cycles are disjoint, (they do not share components), they may be freely ordered as circuits. Figs. 1a, 1b, 1c, and 1d, show the realization of F as given, as well as the realization of the three cycles. All circuits were obtained using the bidirectional MMD+ algorithm [3]. Gates without a control signal are called "simple" and will be assigned a cost of 1. All controlled gates are of the Muthukrishnan-Stroud type, which are active if the control variable has the value 2, otherwise they are inhibited and behave as an identity. In the circuits representations, in analogy with the binary standards, gates controlled by 2 show a black dot, gates controlled by 1 show a grey dot, and gates controlled by 0 show a white dot on the controlling line. Gates controlled by 2 have a cost of 2, whereas gates controlled by 1 or by 0 have a cost of 4 [3], since they need simple gates to shift the controlling signal value to 2 and to recover its original value after leaving the controlled gate. (These simple gates are not shown in the circuits diagrams to simplify the representations, but are considered in the context of post-processing). As basic gates, the set of permutations of {0, 1, 2} is considered, which builds a non-Abelian multiplicative group and is functionally complete. See Table I, where the identity **I** of the group is included for completeness, but obviously no gate is needed.

**Table I**: Basic Ternary Reversible Gates

| Notation | N | $P_{01}$ | $P_{12}$ | X | $X^T$ | (I) |
|---|---|---|---|---|---|---|
| Specification In GF(3) | $2x \oplus 2$ | $2x \oplus 1$ | $2x$ | $x \oplus 2$ | $x \oplus 1$ | $(x \oplus 0)$ |
| Circuit representation | [N] | [01] | [02] | [+2] | [+1] | ---- |

Since the cycles of the example are disjoint, a cascading ordering may be chosen to take advantage of possible merging between pairs of border gates of the cycles. Let G⟨$v_z$⟩ denote a gate G (from Table 1) controlled by a signal with value $v \in \{0, 1, 2\}$ at the line corresponding to $z \in \{x, y\}$. For the above example it becomes apparent that a convenient ordering of the cycles to realize F is (2, 7, 6)(0, 4, 8)(1, 3, 5), because then the last N⟨$2_x$⟩ of the cycle (2, 7, 6) would cancel the first N⟨$2_x$⟩ of the cycle (0, 4, 8). (Compare Figs. 1(c) and 1(b)). Then the realization by cycles has a cost of (10 + 14 − 4) + 20 = 40, whereas the MMD+ realization of F has a cost of 18 (without post-processing) and only 8 gates. The realization by 3-cycles is clearly much more expensive than the direct MMD+ realization.

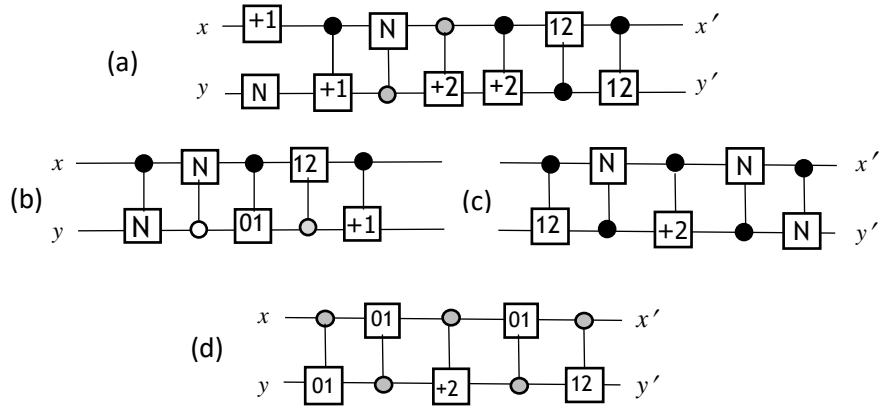

**Fig. 1**: (a) Realization of F with a cost of 18
(b) Realization of the cycle (0, 4, 8) with a cost of 14
(c) Realization of the cycle (2, 7, 6) with a cost of 10
(d) Realization of the cycle (1, 3, 5) with a cost of 20

To continue the decomposition of the 3-cycles into transpositions, we chose the product model, leading to: (0, 4, 8) = (0, 4) (0, 8), (1, 3, 5) = (1, 3) (1, 5), and (2, 7, 6) = (2, 7) (2, 6). The circuits for the transpositions are presented in that order in Fig. 2, showing the possibility of merging neighbor gates if controlled by signals with the same value.

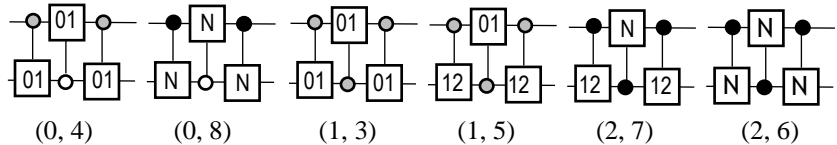

(0, 4)  (0, 8)  (1, 3)  (1, 5)  (2, 7)  (2, 6)

**Fig. 2**: Transpositions derived from the 3-cycles.

From Fig. 2 it is simple to see that the 1$^{st}$, 3$^{rd}$, and 4$^{th}$ transpositions have each a cost of 12; the 2$^{nd}$ has a cost of 8, and the last two have each a cost of 6. Therefore –(if no post-processing is considered)– the realization of the 3-cycle (0, 4, 8) based on transpositions has a cost of 20, the realization of the 3-cycle (1, 3, 5) has a cost of 24, and the realization of the 3-cycle (2, 7, 6) has a cost of 12. All three realizations based on transpositions have a higher cost than the corresponding direct MMD+ realizations.

If post-processing is considered, then it may also be seen in Fig. 2, that when cascading (1, 3) and (1, 5), the gate $P_{01}\langle 1_x \rangle$ of (1, 3) may be merged with the gate

$P_{12}\langle 1_x \rangle$ of (1, 5) leading to a single gate $X\langle 1_x \rangle$. The resulting new circuit is identical with the one in Fig. 1(d) and has obviously the same cost. Similarly when cascading (2, 7) and (2, 6), the gate $P_{12}\langle 2_x \rangle$ of (2, 7) and the gate $N\langle 2_x \rangle$ of (2, 6) may be merged to a gate $X\langle 2_x \rangle$. The resulting circuit is identical with the one shown in Fig. 1(c) and has a cost of 12. No merging is however possible in the circuit for (0, 4)(0, 8), since the neighbor gates have different control values, thus remaining with a cost of 20, which is more expensive than the direct MMD+ realization of (0, 4, 8), which has a cost of 14.

The former analysis gives an explanation to the fact that decomposing permutations down to transpositions will mostly lead to realizations with a higher cost than realizations based on larger cycles. Decomposing a permutation into transpositions may lead to a set of non-disjoint cycles. Therefore, in that case, transpositions cannot be freely ordered. It will depend on whether border gates may be merged or not, to possibly reduce the realizations cost. (The above example illustrates this situation showing that the realization with transpositions is more expensive than the realization based on 3-cycles, because in one of the transpositions no merging post-processing was possible.)

It is fair to point out that there are a few transpositions with low cost. This is the case of transpositions with the structure $(v, (v \oplus w))$ or $((v \oplus w), v)$ where $v \in \{0, 1, 2\}$, $w \in \{1, 2\}$, and $\oplus$ denotes the sum mod 3. If given a global permutation, which decomposition into transpositions leads to "many" low cost transpositions, this decomposition may lead to a realization with a lower cost than a MMD+ based realization. As shown in [2], this happens in only 1.9% of the 9! benchmark functions. Recall that the MMD+ algorithm is greedy and always tries to optimize the prevailing iteration step, but this does not necessarily optimize the final circuit.

However, even in most cases of decompositions leading to low cost transpositions, applying MMD+ to the undecomposed original permutation leads to an equivalent circuit with the same cost or a better realization.

Let F = (0, 7, 1, 4, 3, 8, 6, 2, 5) = (1, 7, 2)(3, 4)(5, 8)(6) = (1, 7)(1, 2)(3, 4)(5, 8)

Each of the obtained transpositions is realizable with a single controlled gate: (1, 7) → $N\langle 1_y \rangle$; (1, 2) → $P_{12}\langle 0_x \rangle$; (3, 4) → $P_{01}\langle 1_x \rangle$; and (5, 8) → $P_{12}\langle 2_y \rangle$. The transpositions have a cost of 4, 4, 4, and 2, respectively. A transpositions based circuit with a cost of 14 is shown in Fig. 3(a). Fig. 3(b) shows one (of several) circuits obtained with MMD+, which comprises the same gates as the circuit based on the product of cycles, except that the gate $P_{12}$ controlled by 2 stands now as the first gate of the circuit and from the straight circuit point of view, this displacement of that gate is not "obvious". However, recall that this gate realizes the transposition cycle (5, 8), which being disjoint from the other cycles, may be moved to any other position, whereas the pair of gates $N\langle 1_y \rangle$ and $P_{12}\langle 0_x \rangle$ should not exchange their positions, because they would realize the 3-cycle (1, 2, 7), which does not belong to the decomposition of F. Furthermore, notice that F could also be decomposed as (2, 1, 7)(3, 4)(5, 8) = (2, 1)(2, 7)(3, 4) (5, 8). However, the transposition (2, 7), i.e. 02 → 21, is no longer a "one gate subcircuit" with a cost of 4. It is realized as $P_{12}\langle 2x \rangle$ $N\langle 2_y \rangle$ $P_{12}\langle 2_x \rangle$ with cost of 6, leading to the full circuit with a cost of 16, (and 6 gates), as shown in Fig 4. MMD+ however, being independent of the cycle decomposition obtains "again" the circuit of Fig. 3(b) with a cost of 14 (and only 4 gates).

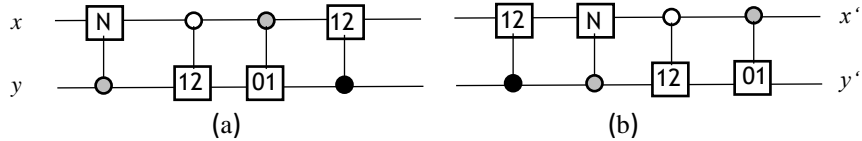

**Fig. 3**: (a) Realization of F as a product of transpositions
(b) MMD+ realization of F.
Both circuits have a cost of 14

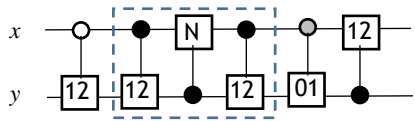

**Fig. 4**: Realization of F as (2, 1)(2, 7)(3, 4)(5, 8) with a cost of 16
The dashed box shows the realization of (2, 7)

## 3   Context dependent post-processing

For exhaustive tests on a large database to have results with statistical significance, post-processing may be considered if the representative patterns may be recognized "almost on the fly". This is e.g. the case of two different neighbor gates controlled by 1 and by 0, since the two simple gates, needed in between on the controlling line, which are neighbors, may be merged, thus reducing the cost of the circuit by 1.

A totally different situation appears when facing the problem of designing a reversible circuit for *one* given function. Different alternatives may be compared within a reasonable time. For example, recall that in the MMD+ algorithm there are cases where a random choice may be taken. If, instead, all alternatives are tested, an optimal solution might be obtained. In the case of cycles, a cyclic reordering may lead to a lower cost of decomposition with transpositions, e.g. (1 3 5 7 ) = (1 3) (1 5)(1 7) has a cost of 28, where as (7 1 3 5) = (7 1)(7 3)(7 5) has a cost of 18. (The inverse situation was illustrated earlier, when an unfortunate cycle representation lead to the circuit of Fig. 4.)

A particular effective post-processing situation, that may be associated to a template [14], is the following: let $G\langle 0_x\rangle$ be followed by $G\langle 1_x\rangle$, (or the other way around), i.e., G is *de facto* controlled by "not 2" on $x$. This sub-circuit has a cost of 8. It may be replaced by a simple G gate on $y$ followed by $G^{-1}\langle 2_x\rangle$ with a total cost of 3. Fig. 5 illustrates the proof. If a gate G is controlled by 0, by 1, and by 2, this is equivalent to a simple G gate on the target line. (Top of Fig. 5). Cascading at both sides with $G^{-1}$ controlled by 2 cancels the $G\langle 2_x\rangle$ gate on the left hand side and leads to the assertion. (Bottom of Fig. 5).

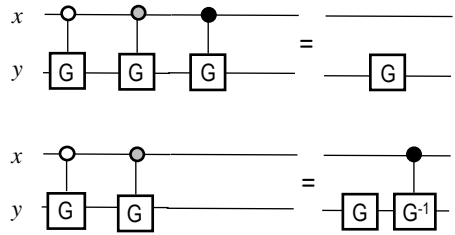

**Fig. 5**: Template to reduce the cost from 8 to 3

A similar, but much weaker improvement may be obtained in the case of $G\langle 0_x\rangle$ followed by $G\langle 2_x\rangle$, with a cost of 6, which may be replaced by a simple G followed by $G^{-1}\langle 1_x\rangle$ with a total cost of 5. The total cost is reduced by 1. The same applies in the case of $G\langle 1_x\rangle$ followed by $G\langle 2_x\rangle$, with a cost of 6, which may be replaced by a simple G followed by $G^{-1}\langle 0_x\rangle$ with a total cost of 5.

## 4   On the class of 2×2-based $n \times n$ reversible circuits

Consider an $n \times n$ reversible circuit with $n-2$ controlling lines and two "working" lines driving a cascade of 2×2 circuits. The controlling lines generate $p^{n-2}$ selecting words, each one of them activating a particular 2×2 reversible circuit, while all others remain inhibited and behave as an identity, connecting the selected circuit to the driving inputs and to the outputs. Fig. 5 shows a simple 3×3 reversible circuit of this class, where $C_0$, $C_1$, and $C_2$ are 2×2 reversible circuits and $c$ denotes the controlling input signal (which is recovered at the output).

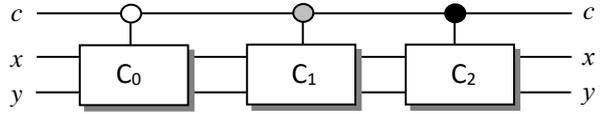

**Fig. 6**: A scheme of a representative of the class of 2×2-based 3×3 circuits.

It is simple to recognize that the circuit of Fig. 6 represents a multiplexer of $C_0$, $C_1$, and $C_2$. Other forms of multiplexing have been used in the past for the synthesis of reversible circuits, as e.g. [4], [9], where generalized gates are generated by multiplexing selected unary functions. The *class* of 2×2-based 3×3 ternary reversible circuits is powerful: it has $(3^2!)^3 = 47,784,725,839,872,000 \approx 47.78 \cdot 10^{15}$ circuits, and the class of 2×2-based 4×4 ternary reversible circuits has $(3^2!)^9 \approx 10.91 \cdot 10^{48}$ circuits, which inherit the main properties of 2×2 circuits: in most cases a direct MMD+ realization will have a lower cost than a cycle based realization. All involved gates will have a high cost, due to the additional control signals, but this will affect both the direct MMD+ realizations and the cycles-based realizations (which are also obtained with MMD+). A natural property of 2×2 ternary reversible circuits is the satisfaction of the near neighbor lines constraint posed by some technologies. This property will occasionally be lost in 2×2-based 3×3 ternary reversible circuits, in the scarce cases where the 2×2 realization includes a simple gate on the target (bottom) line, which in a 2×2-based 3×3 ternary reversible circuit will become controlled by a signal of the upper line. Fig. 7 shows an example of realization the control line for the scheme shown in Fig. 6, including the recovery of the control signal. (An alternative sequence of simple gates to obtain the appropriate control signals could be e.g. N, $X^T$, $P_{12}$).

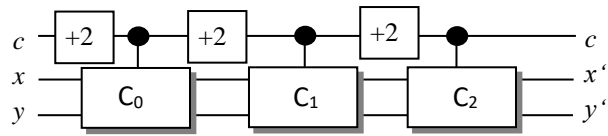

| $c$ | +2 | +2 | +2 | N | $X^T$ | $P_{12}$ |
|---|---|---|---|---|---|---|
| 0 | **2** | 1 | 0 | **2** | 0 | 0 |
| 1 | 0 | **2** | 1 | 1 | **2** | 1 |
| 2 | 1 | 0 | **2** | 0 | 1 | **2** |

**Fig. 7**: Realization of the control line of a 2×2-based 3×3 ternary reversible circuit.

Fig.8 shows a detailed random example of a possible 2×2-based 3×3 ternary reversible circuit. The dotted lines show the additional (selective) control. The chosen circuit also shows that only in the case of a simple gate (on the bottom line) in a 2×2 circuit, (like the gate +2 in $C_2$), when integrated in a 3×3 structure the lines neighborhood may be lost. In all other cases it is preserved.

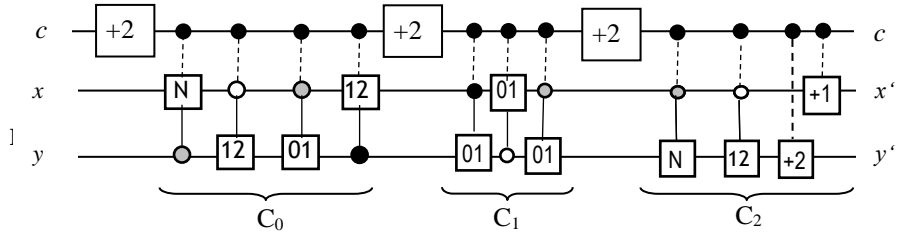

**Fig. 8**: Random example of a 2×2-based 3×3 ternary reversible circuit.

## 5  Conclusions

Based on exhaustive tests, it has been shown [2] that cycle-based synthesis of 2×2 ternary reversible circuits, particularly in the case of transpositions, in almost all cases has a higher cost than realizations obtained with the MMD+ algorithm. The realization of some representative functions was analyzed in detail showing the reasons for the difference in cost. At the same time it was indirectly shown that the MMD transformation algorithm is stronger than it might have been thought of based on [11]. Furthermore we analyzed the 2×2-based 3×3 ternary reversible circuits, we showed that this class of reversible circuits is very large and inherits from 2×2 circuits, the benefits of MMD+ realization as compared with cycle-based realizations. This scalability aspect is valid for all 2×2-based $n \times n$ ternary reversible circuits, since the additional $n - 2$ lines only provide the selectivity for the 2×2 ternary reversible circuits.